\documentclass[fleqn,twoside]{article}
\usepackage{espcrc2}

\usepackage{graphicx}
\usepackage[figuresright]{rotating}

\newcommand{\AmS}{{\protect\the\textfont2
  A\kern-.1667em\lower.5ex\hbox{M}\kern-.125emS}}

\title{The Theory and Simulation of Relativistic Jet Formation: 
Towards a Unified Model For Micro- and Macroquasars}

\author{David L. Meier\address{Jet Propulsion Laboratory, 
        California Institute of Technology, \\
        238-332, 4800 Oak Grove Dr., Pasadena, CA, 91109, U.S.A.}}
       
\begin{document}

\begin{abstract}
I review recent progress in the theory of relativistic jet production, with 
special emphasis on unifying black hole sources of stellar and supermassive 
size.  Observations of both classes of objects, as well as theoretical 
considerations, indicate that such jets may be launched with a spine/sheath 
flow structure, having a much higher Lorentz factor ($\sim 50$) near the axis 
and a lower speed ($\Gamma \sim 10$ or so) away from the axis.  It has become 
clear that one can no longer consider models of accretion flows without also 
considering the production of a jet by that flow.  Furthermore, the rotation 
rate of the black hole also must be taken into account.  It provides a third 
parameter that should break the mass/accretion rate degeneracy and perhaps 
explain why some sources are radio loud and some radio quiet.  

Slow jet acceleration and collimation is expected theoretically, and can explain 
some of the observed features of AGN jet sources.  Finally, relativistic jets 
launched by MHD/ED processes are Poynting flux dominated by nature, and are 
potentially unstable if there is significant entrainment of thermal material.  
\end{abstract}

\maketitle

\section{The Launching of Relativistic Jets}

As this meeting comes only a few months after the meeting on 
microquasars in Cargese, Corsica, my presentation here will be similar 
to that given in Cargese, but with a more general approach to {\em all} 
relativistic jet sources --- Microquasars and Macroquasars alike.  This 
paper, therefore, will be an extension and update of the Cargese paper 
\cite{Meier02}, and the reader will be referred to the latter rather 
frequently.

\subsection{Relativistic Jet Sources and their Speeds}

The first point that I wish to make is that, {\em in attempting to 
understand the launching, acceleration, and collimation of relativistic 
jets, high energy jetted sources of all types should be considered}.  
Their similarities point to closely-related mechanisms and similar 
physics, and their differences give clues on how the general mechanism 
might operate differently under different conditions. 

{\bf Microquasars.}  This class of objects has historically included 
Galactic jet sources, mainly low-mass X-ray binaries (LMXBs) like GRS 
1915+105 and other objects like SS433 (which may be a neutron star). 
Recently, HMXBs like Cyg X-1 have been found to produce jets and, 
therefore, added to the class, and $\gamma$-ray bursts (GRBs) are treated 
as a closely-related object.  However, now that Z and atoll neutron 
star binaries, isolated pulsars, and even core-collapse supernovae 
appear to produce jetted flows, I have suggested that the Microquasar 
class now include {\em all} objects of stellar mass that produce 
relativistic collimated flows.  As discussed in \cite{Meier02}, these 
objects are related not only in their phenomenology and in their 
underlying jet-production mechanism, but also in their common origin 
as the last stages in the evolution of massive stars.

{\bf Macroquasars.}  Like the term 'microquasar', the term 'quasar' 
has evolved --- from the early meaning of quasi-stellar {\em radio} 
source, to encompassing any extragalactic object (radio loud or quiet) 
whose host galaxy is difficult to detect.  To the Macroquasar class 
I also suggest adding the active galactic nuclei (AGN) objects --- 
radio galaxies and Seyfert galaxies --- which are distinguished from 
the others only by the faintness of their central optical source relative 
to the brightness of the surrounding galaxy.  

The recent observations of Blundell \& Rawlings \cite{BR01} are 
extremely important in any attempt to unify all AGN and quasars.  These 
authors found the first Fanaroff\& Riley class I radio quasar, which 
had been known previously as a radio 'quiet' quasar.  Its radio luminosity 
is just below the FR I/FR II break ($\sim 10^{41} {\rm erg~s^{-1}}$), 
or about ten times more powerful than Centaurus A.  While much more 
work needs to be done, the implication is that many, if not all, radio 
'quiet' quasars are actually giant radio galaxies, appearing much like 
Centaurus A, but with a very bright optical core at the center of the 
nucleus.  

While I have suggested that Microquasars be unified on the basis of an 
evolutionary scheme \cite{Meier02}, it is more appropriate to unify 
Macroquasars on the basis of the size, fueling rate, and spin of their 
central black hole (in addition to the jet viewing angle) using, for 
example, theoretical Owen-Ledlow diagrams of the radio-optical plane.  
Figure 1 shows such a scheme, slightly modified from that presented in 
\cite{Meier99}.  Except for the low-mass cutoffs of the FR II objects, 
the lines are {\em computed} from accretion and jet-production 
equations, not simply drawn schematically. Note that this grand unified 
scheme predicts that there are substantial FR I quasars and that their 
class merges into the radio quiet quasar class, as suggested by 
recent observations \cite{BR01}.

\begin{figure*}
\center
\includegraphics[width=15pc,angle=90]{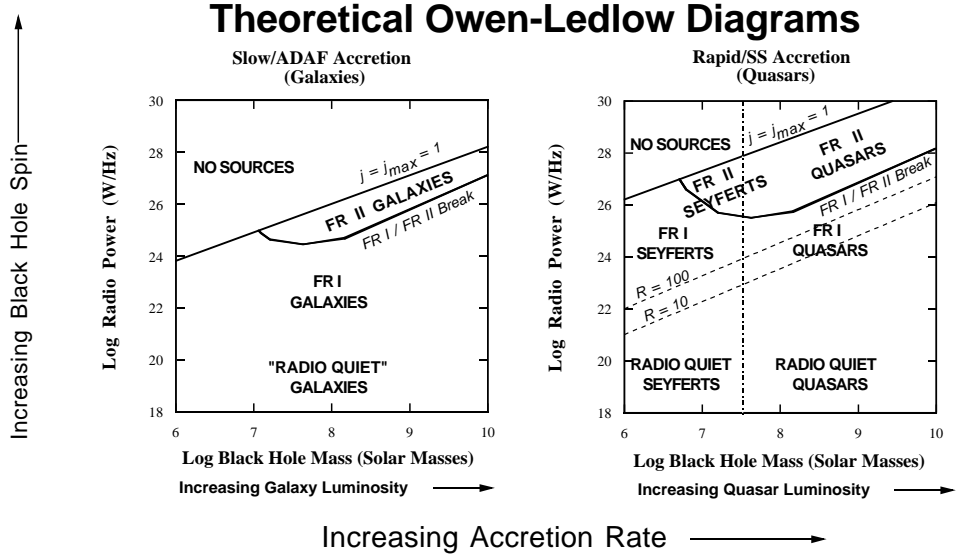}
\caption{Grand unified scheme for Macroquasars, beyond viewing angle 
considerations (after \cite{Meier99}).
Diagrams were {\em computed} from accretion and jet-production equations.
The upper boundary is essentially equation (1) with $\dot{M}/\dot{M}_{Edd} = j = 1$. 
Note the dotted lines of constant radio-to-optical flux ratio $(R = 10, 100)$ and that the 
FR I quasar region occupies the traditional 'radio quiet' quasar region.}
\end{figure*}

{\bf Jet Speeds.}  Measured jet speeds in the above 
sources range from $\sim 0.5~c$ in the neutron star sources, to 
Lorentz factors of $\Gamma \sim 10$ in Micro- and Macroquasar black 
hole systems, to $\Gamma > 100$ in the GRBs.  Current collapsar models 
of long-duration GRBs suggest that the actual jet speed produced by 
the black hole itself may be only of order $\Gamma < 50$, with 
the additional acceleration to $\Gamma$ of several hundred being provided 
by a confining supernova envelope and the subsequent breakout \cite{ZW02}. 
(If this is the case, then one might expect short-duration GRBs to have 
Lorentz factors significantly under 100, as they are thought to be 
associated with neutron star mergers and therefore not occurring inside 
dense envelopes.)  

Broad absorption line (BAL) quasars represent another type of outflow 
that may be related to jets.  Because their P-Cygni lines imply low 
filling factors in the flow, and because of the existence of detached 
absorption troughs, there is some reason for believing that the flow 
may be bi-polar and limb-brightened in nature.  Their velocities 
of $\sim 0.1~c$ are considerably slower than typical jet speeds, and 
radiation pressure may play a key role here in the initial launching. 
Nevertheless, it  is possible that magnetic effects like those 
discussed below are still at work in these outflows, shaping them and 
perhaps providing additional acceleration far from the disk.

\subsection{Of Spines and Sheaths}

The second point that I wish to make is that 
{\em there is some observational evidence that the same source may produce 
jets of rather different Lorentz factors}, either simultaneously or when
the source is in different accretion states.  First of all, there has 
been considerable discussion of the spine-sheath model at this conference 
(see, {\it e.g.}, reference \cite{Giovannini02}).  In addition, I offer 
two other examples.  Intra-day variable (IDV) sources such as PKS 0405-385 
\cite{Rickett02} show evidence of Lorentz factors of up to 75, even when 
the interstellar scintillation model is applied to the variability. When 
this Jansky-level core is de-boosted, one derives only a microjansky-level 
intrinsic flux for this very relativistic flow.  However, for the 
surrounding centimeter flux, with a typical Lorentz factor of 5-10, the 
de-boosted flux is at the millijansky level --- three orders of magnitude
stronger.  The implication is that a great many sources may be producing 
very relativistic 'spine' jet flows that normally are not seen because of 
their weak microjansky-level flux and their highly-beamed nature.

Another clue may be in the behavior and spectrum of Cyg X-1.  Like 
most Microquasars, the source produces a steady jet when it goes into 
the low/hard state \cite{Fender99}.  The non-thermal power-law tail in 
the hard X-ray region in such sources has been suggested to come from 
close to the base of the jet itself \cite{Markoff02}.  In the high/soft 
state, Cyg X-1 does not produce a detectable radio jet, yet the 
$\gamma$-ray spectrum has a power-law tail that extends out to several MeV!  
Clearly, there is still substantial non-thermal activity in what would 
otherwise appear to be a rather thermal and soft source.  While the role 
of this very hard emission is unclear at this point, it is interesting to 
speculate that the $\gamma$-ray tail may indicate the production of a weak, 
but very high Lorentz factor jet at the center of the otherwise thermal disk 
--- a spine for the low/hard state radio jet sheath.  

\subsection{Jet Launching Mechanisms}

As discussed by \cite{Blandford02,Meier_etal01,Meier02} the 
current popular model for launching, accelerating, and collimating 
astrophysical jets is a magnetohydrodynamical/electrodynamical one. 
A strong electromagnetic field in the central engine, coupled with 
differential rotation, serves to convert rotational kinetic 
energy into kinetic energy of outflow.  A magnetic pressure gradient 
(plus, perhaps, the action of thermal and/or radiation pressure) 
lifts the material out of the gravitational potential well, and 
the pinch effect (hoop stress of the magnetic field coiled by the 
rotation) collimates the outflow into a jet.  

The above basic mechanism can be realized in a variety of Micro- 
and Macroquasar situations. For neutron star systems, the 
magnetic field and rotation of the pulsar or protopulsar will 
accelerate plasma trapped in the magnetic field lines.  (This plasma 
can originate from either particle creation in spark gaps or accretion 
flows.)  Ejection of a collimated outflow at roughly the neutron 
star escape speed ($\sim 0.5~c$) provides a natural explanation 
for pulsar and supernova jets, and even possibly SS433-type objects.

In systems with substantial accretion disks the combination of orbital 
motion and a disk coronal magnetic field can provide a similar 
mechanism, first proposed by Blandford and Payne (BP) \cite{BP82}.  Such 
accretion disk MHD winds could operate in both accreting neutron 
star and black hole systems.

Finally, in black hole systems the magnetic field can extract rotational 
energy of the black hole in two different ways.  The first method, 
suggested by Punsly and Coroniti \cite{PC90}, is really an extension 
of the BP mechanism to accretion systems that are significantly 
affected by frame dragging.  Rotation of the space near the black hole, 
if in the same sense as the disk, enhances the disk MHD wind power.  The 
coupling to the black hole rotation is indirect, through material that 
is accelerated into {\em negative energy orbits} inside the ergosphere. 
When accreted by the hole, this material {\em spins down the hole} in a 
magnetic Penrose-type process.  The second method, suggested by Blandford 
and Znajek (BZ) \cite{BZ77}, utilizes direct magnetic coupling with a 
field that threads the horizon and either the accretion disk or an 
outflowing wind, much like the structure of a pulsar wind.

The third point I would like to make, then, is that {\em there are natural 
theoretical reasons for believing that more than one MHD jet launching 
mechanism may be at work in Micro- and Macroquasars}, and that there 
are definite candidates in the different cases.  The following 
identifications are suggested, although the set is certainly subject 
to change as more is learned about these sources.  BP-type outflows 
may be responsible for the lower velocity ($\sim 0.1~c$) outflows 
in black hole systems, shaping them if not also accelerating them. 
The PC/BP mechanism inside the ergosphere may be responsible for 
most jets we see in AGN, quasars, and classical microquasars.  Lorentz 
factors of 3 have been achieved in simulations of this process 
\cite{Koide99}, and values of 10 or more are not unexpected from a 
region where the metric 'rotational velocity' is formally greater 
than $c$.  Finally, it is tempting to identify the very high Lorentz 
factor ($~50$) implied for the IDV spine and for the central 
GRB engine with the BZ mechanism that couples to the black hole horizon 
itself.  Normally expected to generate only a fraction of the 
energy output of the other disk mechanisms \cite{Li00}, the BZ process 
nevertheless could {\em appear} to dominate in observations where 
beaming angles are extremely small.  

The identification of mechanisms like the PC/BP and BZ ones as being 
responsible for most of the AGN jets observed brings up an important 
fourth point.  Black holes with similar mass and accretion rate can 
differ in radio power by several orders of magnitude.  {\em Jet production 
mechanisms that depend on extraction of black hole rotational energy, 
therefore, provide a third parameter that lifts the mass/accretion rate 
degeneracy and potentially can explain why some sources are radio 
loud and some are radio quiet} \cite{WilsonColbert95,Meier99}.  Such 
mechanisms have jet powers that vary significantly with the normalized 
black hole spin $j \equiv J/(GM^2/c)$ but still vary linearly with 
the accretion rate and mass (see Figure 1)

\begin{equation}
L_{jet} = L_{Edd}~~(\dot{M}/\dot{M}_{Edd})~~j^2
\end{equation}

\subsection{The Important Role of Accretion and a Toy Model}

Because the type of jet produced in black hole systems appears related 
to the structure of the accretion flow (low/hard, high/soft, {\it etc.}), 
this brings me to my fifth point.  {\em It is no longer reasonable to 
consider accretion models without also considering jet production}. 
Unfortunately, none of the current accretion models address jet 
production in any meaningful way.  This point is emphasized by 
the association of jet production with the presence of quasi-periodic 
oscillations in the X-ray light and with 'dips' in the X-ray 
emission at essentially the same time as the jet is ejected.  The 
latter occurs not only in Microquasars like GRS 1915+105 but 
also in Macroquasars like 3C 120 \cite{Marscher_etal02}.  In short, 
it is not clear that we have an adequate accretion model that 
will begin to address one of the more important aspects of all 
accreting sources --- jets.  

I therefore propose the following toy scenario, whose main purpose 
is to stimulate further thinking along these lines.  The sheath is 
produced in the accretion disk in low $\dot{M}$ states ({\it i.e.}, 
low/hard state and in X-ray dips in the very high/unstable state).  
It can have Lorentz factors up to 10 or more, so it is produced 
probably near the ergosphere.  It dominates in low-luminosity 
sources (FR Is, low-luminosity AGN, and persistent X-ray binary jets). 
The low/hard power-law tail in Cyg X-1 may originate in the sheath.

The spine, on the other hand, is produced only in high $\dot{M}$ states, 
when rapid accretion can press the magnetic field onto the black hole 
(high/soft state and between dips in the very high/unstable state). It 
can have Lorentz factors up to 50 and higher, so it is produced probably 
very near the horizon.  The spine is more important in high-luminosity 
sources (FR IIs, GRBs, and high-luminosity X-ray transients).  In Cyg 
X-1 the hard MeV tail in the high/soft state may originate in this jet. 

\section{Jet Collimation, Acceleration, and Stability}

\subsection{Collimation}

As pointed out in \cite{Meier02}, there are both theoretical 
and observational reasons for concluding that slow acceleration 
and collimation is probably the norm for jet outflows in these 
sources.  Non-relativistic \cite{KLB99} and relativistic \cite{VK01} 
models of MHD wind outflows attain solutions where the wind 
opening angle is wide near the accretion disk and then narrows 
slowly over several orders of magnitude in distance from the disk. 
Furthermore, recent observations of M87, for example, by \cite{JBL99} 
suggest that the opening angle of the jet is more than $60\deg$ at 
the base, collimating to a few degrees only after a few hundred 
Schwarzschild radii.  Furthermore, the lack of significant 'Sikora' 
bump in the X-ray light of most radio quasars indicates that the 
flow at the base of most quasar jets must also be broad and probably 
sub-relativistic, only accelerating to relativistic flow much further 
from the black hole.  

These observational results have important implications for the 
spine/sheath model.  If there is, indeed, a BZ-type high-$\Gamma$
jet produced near the black hole, then this jet 
nevertheless cannot dominate in even the most powerful of radio 
quasars.  Such a flow would intercept the soft photons from the 
accretion disk, Compton-scatter them to hard X-ray energies, and 
produce a substantial Sikora bump.  The lower Lorentz factor flow, 
collimated and accelerated slowly, must still dominate in even the 
brightest of radio quasars.  This conclusion is consistent with the 
observation that the parsec-scale jets in both FR I and FR II radio 
sources appear to have similar structures and speeds 
\cite{Giovannini02}.

\subsection{Attaining Relativistic Speeds: Poynting Flux-Dominated Jets}

It is important to realize that, if an MHD/ED mechanism for jet 
acceleration is adopted, then this implies that (at least initially) 
the jets so-produced must be Poynting flux-dominated (PFD).  By definition, 
$\Gamma >>1$ implies that the kinetic energy greatly exceeds the rest 
mass-energy of the flow.  And, for an MHD jet, the final velocity is 
at least of order the Alfven speed, so the Alfven Lorentz factor also 
must be large:  $\Gamma_{A} = V_{A}/c = B/(4 \pi \rho c^2)^{1/2} >> 1$.
That is, the field lines must have low mass-loading, and the energy 
flow must be dominated by the flow of electromagnetic energy (Poynting 
flux), not kinetic energy.  As the flow accelerates, Poynting flux 
is slowly converted into kinetic energy flux, until the two are of the 
same order of magnitude \cite{VK01,V02}.  Eventually, mass entrainment 
from the interstellar medium can increase the baryon loading, 
decreasing the Poynting flux domination.  

\subsection{Stability of Poynting Flows}

So far no fully relativistic numerical simulations of PFD 
flows have been performed.  The best results to date are from 
three-dimensional {\em non-relativistic} simulations \cite{Naka01,NM03}. 
We find that the stability is critically dependent on how severe the 
mass entrainment in the jet is --- specifically on the {\em gradient} of 
the plasma parameter $\beta \equiv p_{gas}/(B^2/8\pi)$.  (In the following, 
remember that $\beta$ is always less than unity for PFD jets, if the plasma 
is reasonably cold $P \leq \rho c^2$.)  If $\beta$ decreases or remains 
small as the jet propagates outward (mass loading becomes even less or 
stays the same), then we find that the PFD jet remains stable.  However, 
if $\beta$ {\em increases} (entrains significantly more thermal material), then we find 
that the jet is likely to be unstable to the helical kink instability, {\em 
even if the jet still remains magnetically dominated throughout the simulation}. 
Apparently even a small amount of pressure in the flow builds up over large 
distances, triggering a helical kink and, therefore, turbulence in the jet. 

We tentatively suggest that this may be part of the reason why FR II sources 
appear only in elliptical galaxies.  The presence of a high gas density in 
spiral galaxies may allow the entrainment of relatively more matter than in 
ellipticals, especially if the propagation direction of the jet makes a 
small angle to the plane of the spiral disk.  While active ellipticals do 
have a substantial amount of gas, that material will generally be in a disk 
about the active nucleus with a rotation axis more or less aligned with the 
central black hole and, therefore, the jet.  

\section{Conclusions}

In the review above I have emphasized several important points 
in the study of relativistic jets:
\begin{enumerate}
\item{High energy jet sources of all types should be considered 
when attempting to understand relativistic jets.  Micro- and 
Macroquasars both provide important clues to the mechanisms at 
work.}
\item{There are observational reasons for believing that the same 
source may produce jets of rather different Lorentz factors, either 
simultaneously or in different accretion states.  This may lead to a 
spine/sheath jet structure.}
\item{Similarly, there are natural theoretical reasons for believing 
that more than one MHD/ED jet launching mechanism may be at work in 
a give black hole engine, and there are candidates for both spine 
and sheath.} 
\item{Some jet production mechanisms at work near the black hole
rely on the extraction of black hole rotational energy.  This 
provides a third parameter, in addition to black hole mass and 
accretion rate, that potentially can explain why sources with 
similar optical appearance are radio loud and some are radio quiet.}
\item{It is no longer reasonable to consider accretion 
models without also considering jet production as an integral part 
of the accretion process.  Many, if not all, sources produce jets, and 
it is clear that the production of a jet is affected by, and can affect, 
the structure of the accretion flow.}
\item{Slow acceleration and collimation appears to be the norm, both 
from observational and theoretical investigations.}
\item{By nature, relativistic jets that are launched via MHD/ED processes 
will be Poynting-flux dominated (PFD).}
\item{PFD jets remain stable as long as they do not entrain a significant 
amount of thermal material, or if they become even more magnetically dominated.  
However, if there is a significant amount of entrainment of hot plasma, 
such that the plasma $\beta$ has a positive gradient, then the jet is 
may be subject to the helical kink instability.}
\end{enumerate}


\begin{thebibliography}{9}
\bibitem{Blandford02} R.Blandford, this volume (2002).
\bibitem{BP82}    R.Blandford and D.Payne, MNRAS, 199, 883 (1982).
\bibitem{BZ77}    R.Blandford and R.Znajek, MNRAS, 179, 433 (1977).
\bibitem{BR01}    K.Blundell and S.Rawlings, Astrophys. J., 562, L5 (2001).
\bibitem{Fender99} R.Fender, in Black Holes in binaries and galactic nuclei, 
                  ed. L. Kaper et al., Springer-Verlag (1999).
\bibitem{Giovannini02} G.Giovannini, this volume (2002).
\bibitem{JBL99}   W.Junor et al., Nature, 401, 891 (1999).
\bibitem{Koide99} S.Koide et al., in Proc. of 19th Texas Symposium on 
                  Relativistic Astrophysics, ed. E.Auborg et al., 
                  World Scientific (1999).
\bibitem{KLB99}   R. Krasnopolsky et al., Astrophys.J., 526, 631 (1999).
\bibitem{Li00}    L.-X. Li, Phys Rev D, 61, 084016 (2000).
\bibitem{Markoff02} S.Markoff et al., astro-ph/0210439 (2002).
\bibitem{Marscher_etal02} A.Marscher et al., Nature, 417, 625 (2002).
\bibitem{Meier99} D.L. Meier, Astrophys. J., 522, 753 (1999).
\bibitem{Meier_etal01} D.L. Meier et al., Science, 291, 84 (2001).
\bibitem{Meier02} D.L. Meier, in 4th Microquasar Workshop, in press.
\bibitem{Naka01}  M.Nakamura et al., NA, 6, 61 (2001).
\bibitem{NM03}    M.Nakamura and D.L.Meier, in preparation.
\bibitem{PC90}    B.Punsly and F.Coroniti, Astrophys. J., 354, 583 (1990).
\bibitem{Rickett02} B.Rickett, L.Kedziora-Chudczer, and D.Jauncey, Astrophys.J., 
                  581, 103 (2002).
\bibitem{VK01}    N.Vlahakis and A.Konigl, Astrophys.J., 563, L129 (2001).
\bibitem{V02}     N.Vlahakis, this volume (2002).
\bibitem{WilsonColbert95} A.S.Wilson and E.J.M.Colbert, Astrophys.J., 
		  438, 62 (1995).
\bibitem{ZW02}    W.Zhang and S. Woosley, astro-ph/0209482 (2002).
\end{thebibliography}
\end{document}